# Edge conduction in monolayer WTe$_2$

Zaiyao Fei[1], Tauno Palomaki[1], Sanfeng Wu[1], Wenjin Zhao[1], Xinghan Cai[1], Bosong Sun[1], Paul Nguyen[1], Joseph Finney[1], Xiaodong Xu[1,2*], and David H. Cobden[1*]

[1]Department of Physics, University of Washington, Seattle WA 98195, USA
[2]Department of Materials Science and Engineering, University of Washington, Seattle, WA
*Corresponding authors: cobden@uw.edu, xuxd@uw.edu

**A two-dimensional topological insulator (2DTI) is guaranteed to have a helical 1D edge mode[1-11] in which spin is locked to momentum, producing the quantum spin Hall effect and prohibiting elastic backscattering at zero magnetic field. No monolayer material has yet been shown to be a 2DTI, but recently the Weyl semimetal WTe$_2$ was predicted[12] to become a 2DTI in monolayer form if a bulk gap opens. Here, we report that at temperatures below about 100 K monolayer WTe$_2$ does become insulating in its interior, while the edges still conduct. The edge conduction is strongly suppressed by in-plane magnetic field and is independent of gate voltage, save for mesoscopic fluctuations that grow on cooling due to a zero-bias anomaly which reduces the linear-response conductance. Bilayer WTe$_2$ also becomes insulating at low temperatures but does not show edge conduction. Many of these observations are consistent with monolayer WTe$_2$ being a 2DTI. However, the low temperature edge conductance, for contacts spacings down to 150 nm, is below the quantized value, at odds with the prediction that elastic scattering is completely absent in the helical edge.**

Experimental work on 2DTIs to date has focused on quantum wells in Hg/CdTe[4-7] and InAs/GaSb[9-11] designed to achieve an inverted band gap. These heterostructures show edge conduction as anticipated[13,14], but they also present some puzzles. One is that the conductance at low temperatures is not perfectly quantized, becoming small in long edges[13] and showing mesoscopic fluctuations as a function of gate voltage[5,7,10]. This is inconsistent with the predicted absence of elastic backscattering at zero magnetic field, although several possible explanations have been put forward for the discrepancy[15-20]. Another is that the edges show signs of conducting even at high magnetic field[21,22], contrary to expectations that helical modes, protected by time-reversal (TR) symmetry at zero field, should Anderson-localize once this symmetry is broken. An additional complication is that non-helical edge conduction may also be present, due for instance to band bending when a gate voltage is applied[23].

Identification of a natural monolayer 2DTI, which lacked some of these discrepancies and which could be probed, manipulated, and coupled with other materials more easily than quantum wells, would be helpful for elucidating and exploiting TI physics. Band structure calculations predict that certain monolayer materials are intrinsically topologically nontrivial[12]. An example is monolayer WTe$_2$, which has the T′ structure illustrated in Fig. 1a. Three-dimensional WTe$_2$, in which such monolayers are stacked in the orthorhombic T$_d$ structure, has recently attracted attention as a type-II Weyl semimetal[24,25] that exhibits extreme non-saturating magnetoresistance[26,27] related to the closely balanced electron and hole densities[28-30]. Calculations suggest that the monolayer will be likewise a semimetal[12,30], its Fermi surface comprising two electron pockets (green) and one hole pocket (gray) as shown in Fig. 1b, with areal densities $n = p \approx 1.6 \times 10^{13}$ cm$^{-2}$. If this is correct then the helical edge modes are always degenerate with bulk states (Fig. 1c). In contrast, in a 2DTI the edge modes span a band gap and cannot be mixed by TR-invariant perturbations, so that they dominate transport when the Fermi energy $E_F$ is in the gap. Here we will present evidence that at low temperatures monolayer WTe$_2$ exhibits an insulating bulk state and edge conduction, and describe the properties of the edge conduction, including its dependence on gate voltage, magnetic field, temperature, contact separation, and bias. We will then compare the behavior with that expected for helical modes in the presence of disorder expected to be present at the monolayer edge.

To make devices, WTe$_2$ sheets exfoliated from flux-grown crystals[27] were fully encapsulated in thin (~10 nm) hexagonal boron nitride in a glove box to prevent degradation from exposure to air[31]. Each device had thin (~5 nm) Pt or Pd contacts patterned on the hBN beneath the WTe$_2$ and a top few-layer-graphene gate (see Supplementary Information 1 for details). Figures 1f-h show representative two-terminal measurements of the differential conductance $G_{diff}$ of a trilayer, a bilayer, and a monolayer device. Each of these devices has a row of contacts along one edge of the WTe$_2$ sheet, as visible in the optical micrograph of monolayer device MW1 in Fig. 1d. For these particular measurements two contacts and the gate were connected as shown in Fig. 1e and a small (3 mV) dc bias was superposed on the 100 μV ac excitation. This dc bias affects only the lowest temperature measurement (1.6 K), by suppressing a zero-bias anomaly (ZBA), as will be explained later. The conductance measurements shown in Figs. 2 and 3 and in Supplementary Information 6 are made with no dc bias, i.e., in linear response.

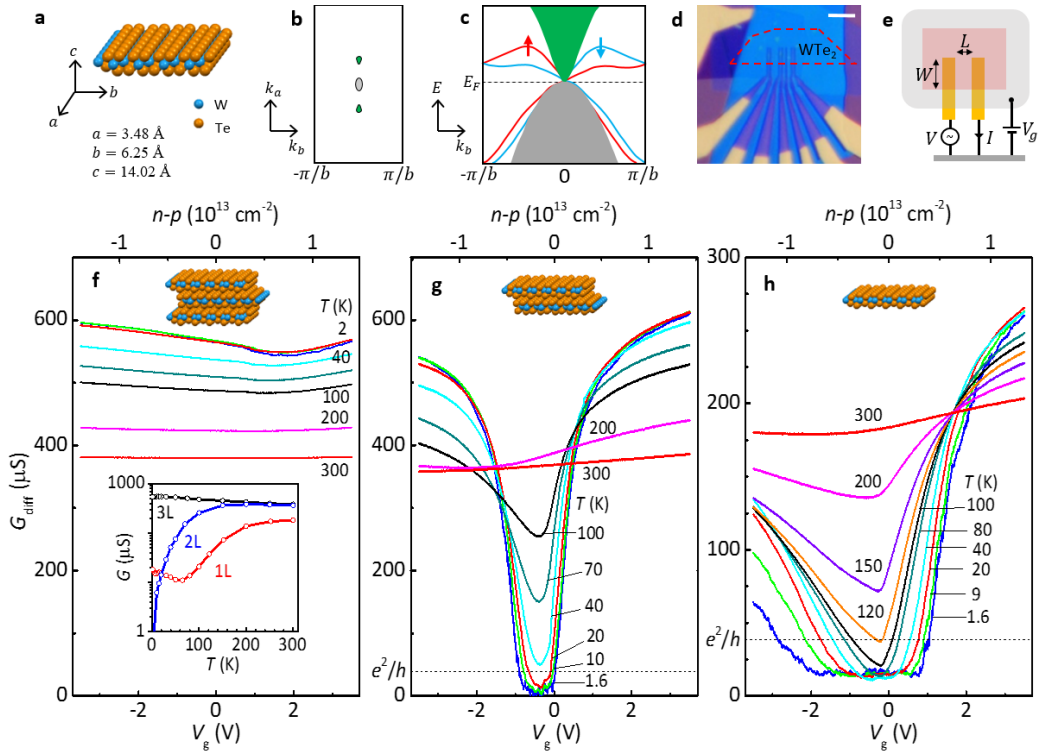

**Figure 1 | Two-terminal characteristics of WTe$_2$ devices. a,** Structure and lattice constants of monolayer WTe$_2$. Tungsten atoms form zigzag chains along the a-axis. **b,** Sketch of its calculated Fermi surface, showing two electron (green) and one hole (gray) pockets. **c,** Sketch of calculated bands in a strip of monolayer WTe$_2$. The spin-resolved helical edge modes (red and blue lines) are predicted to be degenerate with bulk states (gray and green) at all energies. **d,** Optical image of monolayer WTe$_2$ device MW1. Scale bar, 5 μm. **e,** Schematic two-terminal measurement configuration, indicating also the voltage applied to the few-layer graphene top gate (gray). The pink region is the monolayer WTe$_2$. **f-h,** Temperature dependence of the characteristics for similar contact pairs on a trilayer ($L$=0.20 μm, $W$=3.4 μm), bilayer ($L$=0.26 μm, $W$=3.1 μm), and monolayer ($L$=0.24 μm, $W$=3.3 μm) device (MW1) respectively. Here the differential conductance $G_{diff}$ is measured with a small (3 mV) dc bias to suppress effects of a zero-bias anomaly in the 1.6 K sweep for the monolayer. The inset to **f** compares the temperature dependence of the conductance minimum in the three cases.

On cooling from 300 K the trilayer (Fig. 1f) shows metallic behavior at all $V_g$, the conductance rising steadily before saturating at the lowest temperatures, consistent with the behavior[26] of bulk WTe$_2$. The bilayer (Fig. 1g) develops a strong $V_g$ dependence with a sharp minimum near $V_g = 0$, while remaining metallic at large $V_g$. The minimum drops steadily and below ~20 K it broadens and reproducible mesoscopic fluctuations appear. The monolayer (Fig. 1h) first develops a similar but wider minimum, but below ~100 K the minimum stops dropping and instead broadens into a plateau of conductance, here at ~16 μS, on which there are mesoscopic fluctuations. The inset to Fig. 1f compares the temperature dependence in the three cases. We will show below that the plateau seen only in the monolayer is due to edge conduction remaining when the bulk becomes insulating below ~100 K.

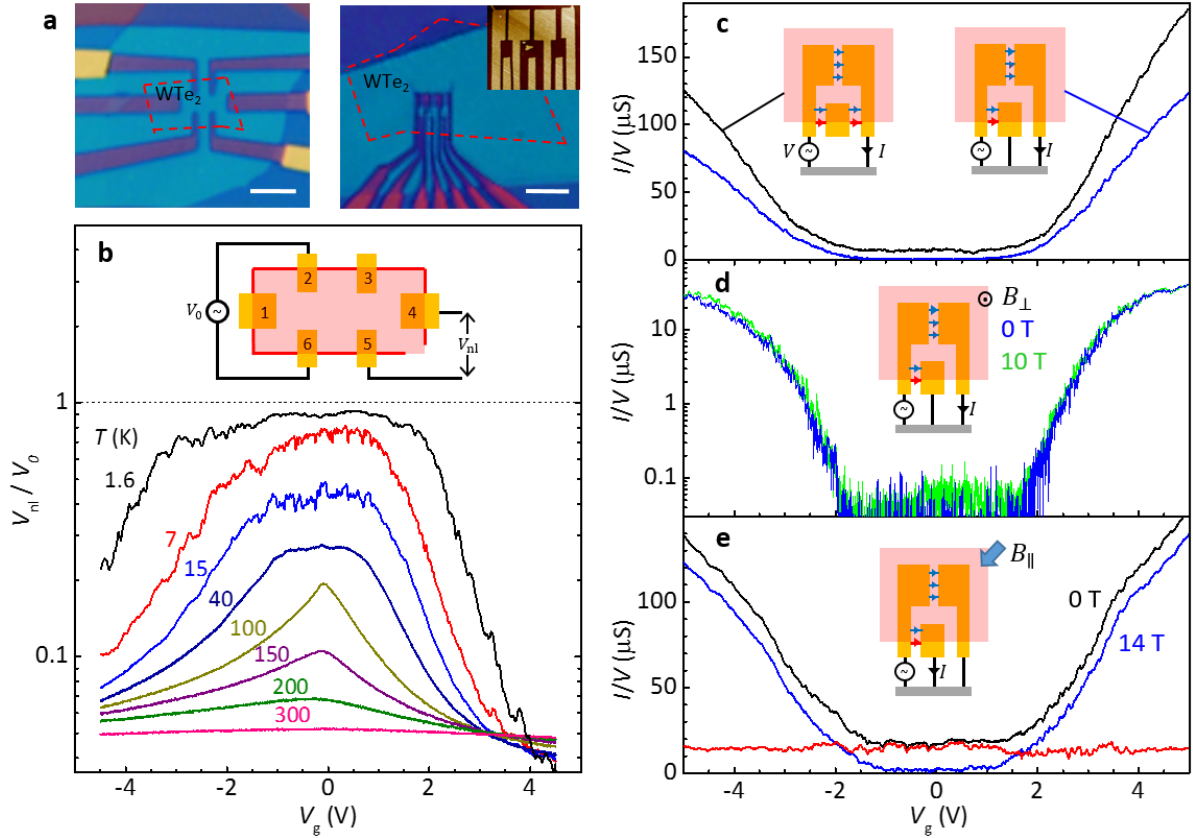

**Figure 2 | Distinguishing edge and bulk conduction. a,** Optical images of the electrodes for monolayer devices MW2 (left) with a Hall-bar layout and MW3 (right) designed to distinguish edge and bulk contributions. Red dashed lines show positions of the monolayer sheets transferred onto them. Scale bars are 5 μm. Inset: AFM image of the electrodes. **b,** Nonlocal measurements on device MW2 in the configuration shown in the inset ($V_0 = 100$ μV at 11.3 Hz). **c-e,** Measurements on MW3 (the contact separation along the edge is 0.22 μm; the pincer spacing is 0.28 μm). **c,** As indicated by the insets, the black trace is the two-terminal conductance between the outer contacts and the blue trace is $I/V$ measured with the center contact grounded ($T = 10$ K). In the insets, red and blue arrows indicate edge and bulk current paths, respectively. **d,** Same measurement as for the blue trace in **c** but at 1.6 K, on a logarithmic scale, showing the very weak effect of a perpendicular magnetic field of 10 T. **e,** Effect of in-plane magnetic field $B_\parallel = 14$ T on $I/V$ between adjacent contacts ($T = 10$ K). The red trace is the magnitude of the decrease.

Edge conduction is normally detected using nonlocal measurements[6] such as those shown in Fig. 2b. Here we apply a small excitation $V_0$ between contacts 2 and 6 on opposite edges of monolayer device MW2 (Fig. 2a, left image) that has approximate Hall-bar geometry, and we detect the nonlocal voltage $V_{nl}$ induced between contacts 4 and 5 which are far out of the normal current path. At low $T$ and small $V_g$, $V_{nl}/V_0$ grows large, suggesting that in this regime most of the current follows the edge. At higher $T$ or larger $V_g$, $V_{nl}/V_0$ falls off as more current takes the direct path through the bulk. Detailed studies showed that the reason $V_{nl}/V_0$ approaches unity at low $T$ is that in this particular device the conductance of edge 4-5 is suppressed more than that of the others by a ZBA (see discussion of Fig. 4), so that contact 4 becomes effectively connected only to contact 2, and 5 to 6.

Although this measurement indicates that the current follows the edge, the shape of the WTe₂ flake in device MW2 is unsuitable for quantitative separation of bulk and edge contributions. To address this, we designed monolayer device MW3 (Fig. 2a, right image) which employs a series of alternating pincer-shaped contacts overlapping one straight edge of a monolayer flake, as shown schematically in the insets to Fig. 2c. The two-terminal linear conductance between a pair of pincers (black trace in Fig. 2c), here at 10 K, behaves similarly to a pair of adjacent contacts in device MW1 (Fig. 1h). However, if the smaller rectangular contact interposed between them is grounded (blue trace in Fig. 2c), so that any current flow near the edge is shorted out, $I/V$ is suppressed nearly to zero around $V_g = 0$. This confirms that around $V_g = 0$ most of the current flows near the edge. However, for $V_g$ larger than about ±2 V conduction does occur through the 2D bulk, directly across the gap between the pincers. Fig. 2d shows measurements of the same quantity at 1.6 K, on a logarithmic scale, both with (green trace) and without (blue trace) a perpendicular magnetic field $B_\perp = 10$ T. Near $V_g = 0$ at this temperature the bulk conductance is unmeasurably small. Below ~100 K it is approximately activated, while above it rises roughly linearly with $T$ (see Supplementary Information 5). The effect of the perpendicular magnetic field is clearly very small; the same is true for an in-plane magnetic field.

In Fig. 2e the black trace is a measurement at zero magnetic field between two adjacent contacts, using the configuration shown in the inset where the rightmost contact (not shown) is grounded to eliminate current along paths not directly between the adjacent contacts. The blue trace is the same measurement done with an in-plane field $B_\parallel$ of 14 T. Since the bulk conductivity is almost immune to magnetic field the decrease in $I/V$ must be associated with the edge. Near $V_g = 0$, where the bulk is insulating, $I/V$ drops nearly to zero, implying that the edge conduction is strongly suppressed by the magnetic field. In addition, the magnitude of the drop, plotted in red, is similar at all $V_g$. This implies that the edge makes a roughly constant contribution to the conductance, independent of gate voltage and bulk conductivity.

The edge conduction can be isolated by working at $T$ and $V_g$ low enough that bulk conduction is negligible, corresponding for example to the plateau region in Fig. 1h. In this regime we find that the section of edge between each pair of adjacent contacts behaves as an independent two-terminal conductor. This is demonstrated by the effect of grounding the central contact shown in Fig. 2c, and also the fact that 2- and 4-terminal measurements on an edge in this regime give identical results (Supplementary Information 3). Conduction along the other edge, passing around the entire flake, is negligible, as we checked by grounding a third contact and seeing that it had no measurable effect. Figure 3 shows the effects of magnetic field and temperature on the linear-response conductance $G_{edge}$ for one edge in device MW2. Figure 3a shows the $T$ dependence at zero field. On cooling from 50 K to 10 K, $G_{edge}$ increases, but below ~10 K at most $V_g$ it decreases again as the mesoscopic oscillations grow. The inset shows the $T$ dependence at a particular $V_g$ where $G_{edge}$ stays level below 10 K. Figure 3b shows the effect of $B_\parallel$, oriented as shown in the inset to Fig. 3c, at 1.6 K. Figure 3c shows $B_\parallel$ sweeps at $V_g = 0$ for a series of temperatures. At moderate $B_\parallel$ and $T$ the behavior is quite well described by the activated function $G_{edge} = G_0 e^{-\alpha B_\parallel/T}$ where $G_0 = 17$ μS and $\alpha = 5$, plotted as the red dashed lines. The effect of a perpendicular field ($B_\perp$) is similar but weaker, as illustrated in the lower inset.

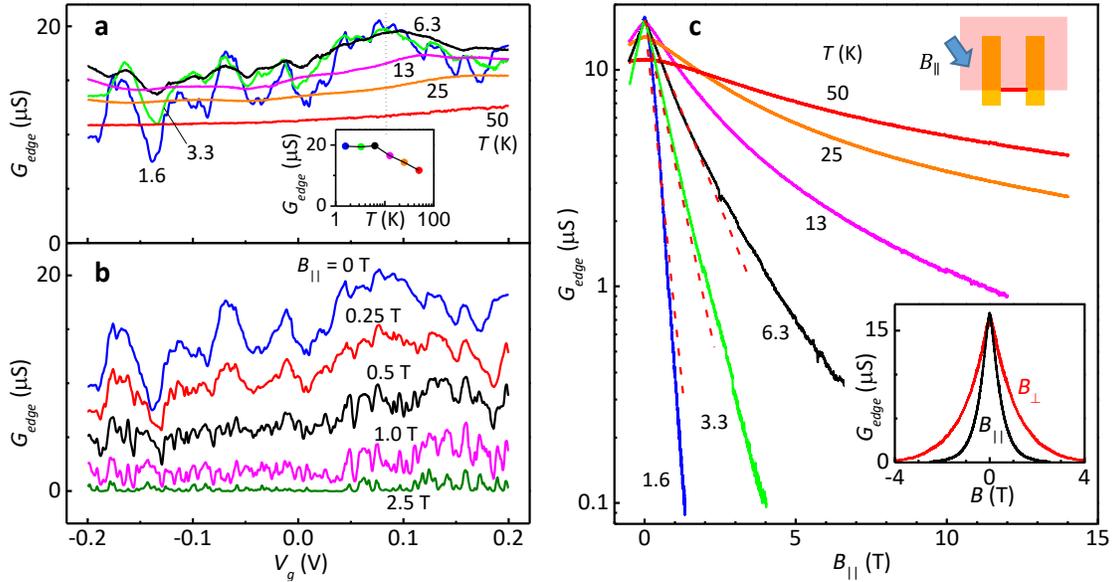

**Figure 3 | Temperature and magnetic field dependence of the edge conductance. a.** Temperature dependence of the conductance at zero magnetic field between adjacent contacts vs gate voltage. To emphasize that the bulk conduction is negligible we label the conductance $G_{edge}$. (Device MW2, $L = 1.3$ μm, $V = 100$ μV ac). Inset: $T$ dependence at $V_g = +0.08$ V. **b.** Effect of in-plane magnetic field $B_\parallel$ at 1.6 K. **c.** Sweeps of $B_\parallel$ at $V_g = 0$ for various temperatures. Dashed lines plot $G_0 \exp(-\alpha B/T)$ with $G_0 = 17$ μS, $\alpha = 5$. Upper inset: orientation of $B_\parallel$ relative to the edge. Lower inset: Comparing perpendicular and in-plane fields at 1.6 K ($V_g = 0$).

The edge conduction is often highly nonlinear at small biases. Figure 4a shows a typical $I-V$ curve at $B_\parallel = 0$ (black) and at 10 T (blue). Figure 4b shows the corresponding differential conductance. At $B_\parallel = 0$ there is a sharp dip in $dI/dV$ at $V = 0$, or zero-bias anomaly[32] (ZBA), whereas at 10 T there is a sharp threshold for current flow. All edges show some dip at 1.6 K, but the strength of the ZBA varies greatly between different edges and as a function of $V_g$. It always deepens as $T$ decreases, as illustrated for one case in Fig. 4c. The mesoscopic fluctuations that grow on cooling are linked to the ZBA, which varies in strength between devices. When a small dc bias is applied to suppress the anomaly the fluctuations are suppressed and we see a flatter edge conduction plateau, which is more representative of the generic behavior of all devices factoring out the effect of the ZBA. This is why in Fig. 1 we plotted $G_{diff} = dI/dV$ at $V = 3$ mV.

In Fig. 4d we compile measurements for 19 adjacent contact pairs in four different monolayer devices, color coded by device, at zero magnetic field. The edge length $L$, which ranges from 0.16 to 5.5 μm, was estimated from atomic force microscope AFM images. For each edge we show the linear-response conductance, averaged over a window of $V_g$ in which the bulk contribution is negligible, at both 10 K (solid circles) and 1.6 K (open circles). The error bars indicate the extreme values of the fluctuations in that window (see Supplementary Information 6). At both temperatures the average conductance tends to decrease with $L$, but the trend is rather weak compared with the large, seemingly random variations. The edges with the weakest $T$ dependence in this range, i.e., the weakest effect of the ZBA, also have the highest conductance, ~20 μS.

We now discuss the compatibility of the above observations with the scenario of a helical edge mode, in comparison with a trivial edge mode or carrier accumulation due to band-bending. First, the monolayer edge conductance is roughly independent of $V_g$, and therefore chemical potential, over the entire accessible range (Fig. 2e). This is consistent with a single gapless mode, and not with carrier accumulation due to band bending, or a trivial edge mode, which would be gate dependent with a gap somewhere. Second, we see no edge conduction in bilayers (Fig. 1g). This can be explained by the fact that TR symmetry does not prohibit backscattering at the bilayer edge if the electron changes layer (the pair of coupled edges is not helical), whereas band-bending effects should be similar to those in a monolayer. Third, the conductance is dramatically suppressed by $B_\parallel$ (Fig. 3c), consistent with the expectation that elastic backscattering is allowed once TR symmetry is broken.

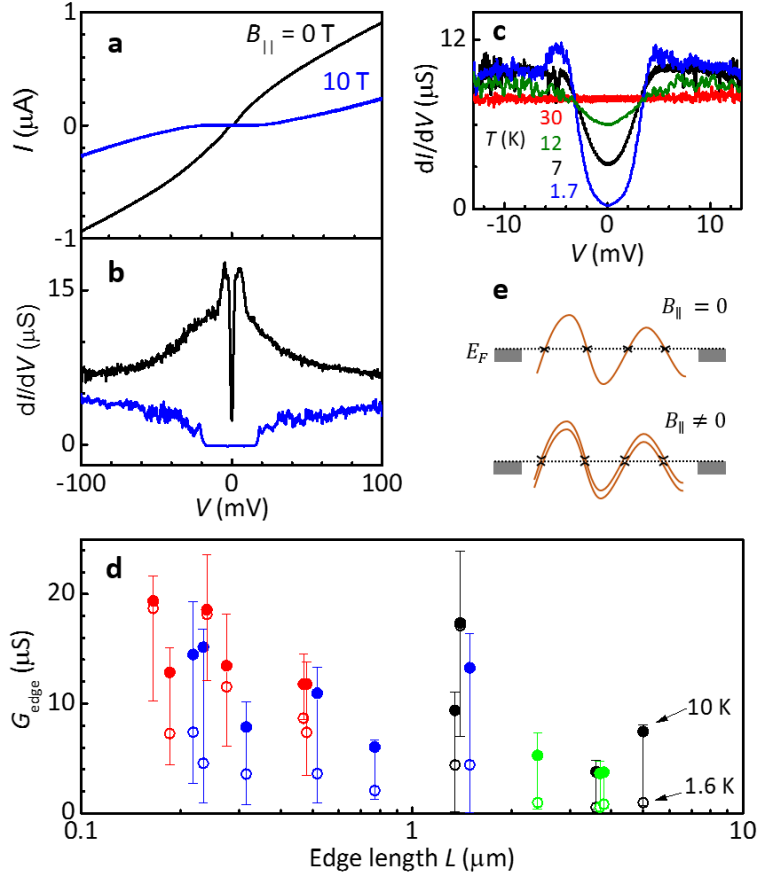

**Figure 4 | Statistics, nonlinear properties, and simple picture of edge conduction. a,** $I - V$ curves at $B_\parallel = 0$ (black) and 10 T (blue) for device MW1, $L = 1.5$ μm ($V_g = 0$, $T = 1.6$ K). **b,** Corresponding differential conductance. A zero-bias anomaly (ZBA) is seen at $B_\parallel = 0$. **c,** Temperature dependence of the ZBA at $V_g = -0.68$ V. **d,** Gate-averaged linear-response edge conductance for 19 adjacent-contact pairs (blue, device MW1; black, MW2; red, MW3; green, MW4) at 10 K (solid circles) and 1.6 K (open circles). Vertical bars show the full range of the mesoscopic fluctuations. **e,** Cartoon picture of combined effects of smooth static disorder and a magnetic field-induced gap along the edge.

It is consistent with a single mode that the linear edge conductance $G_{edge}$ never exceeds the quantized value of $e^2/h = 38.7$ μS expected in the low-temperature limit when elastic backscattering is completely prohibited (Fig. 4d). It is also encouraging that conductance of this order can occur for micron-scale edges in spite of what may be very strong disorder due to the random chemical bonding at the torn edge of the exfoliated monolayer. On the other hand, even when the $T$ dependence is small and for the shortest edges, $G_{edge}$ only reaches about half $e^2/h$ at a peak. One possible factor is imperfect transmission between the metal contacts and the edge. (The requirement of spin relaxation in the metal contacts when spin-polarized current is exchanged with a helical edge might contribute to this). Another is backscattering from multiple magnetic impurities[16-19] or puddles in the disorder potential[20], ideas introduced to explain the deviations from quantization also observed in the quantum well systems. If some form of scattering is allowed at points in a quantum wire it is natural for a ZBA to develop due to interaction effects, such as occurs in a Luttinger liquid[16,33,34]. Nevertheless, quantized conductance would be a definitive signature of a 2DTI and our monolayer WTe2 devices fail in this respect.

We believe the following simple picture provides a helpful framework for the assembled measurements. We posit that there is a single helical edge mode, which follows the physical edge of the monolayer and effectively experiences a large but smooth disorder potential, for example due to trapped charges. As a result, the energy at which the left- and right-going branches are degenerate fluctuates up and down along the edge, passing through $E_F$ at multiple points. This situation is sketched in Fig. 4e. If one accepts that some kind of inter-branch scattering is possible in spite of the helical protection, it is likely to be strongest at these "weak points" where no momentum transfer is required. At $B_\parallel = 0$, in some edges the average linear conductance $G_{edge}$ is not much less than $e^2/h$ and so the scattering must be weak, yet we see large, rapid mesoscopic fluctuations as a function of chemical potential (Fig. 3a). If the origin of these is quantum interference, then since no Feynman paths enclose magnetic flux in a 1D wire we expect no corresponding fluctuations as a function of $B_\perp$, as is the case (see Fig. 3c inset). As $T$ decreases, the scattering from the weak points will strengthen at energies near $E_F$ due to interaction effects[33,34], possibly producing the ZBA. Edges longer than a few hundred nm (see Fig. 4d) are not phase coherent and so have smaller conductance due to classical addition of resistance. Also, as $T$ rises the coherence length will decrease, consistent with the fact that $G_{edge}$ tends to decrease with $T$ when the ZBA is small (see Fig. 3a, above ~6 K).

In this picture, at $B_\parallel \neq 0$ a gap $\Delta = g\mu_B B_\parallel$ opens in the helical mode, where $\mu_B$ is the Bohr magneton and $g$ is an effective g-factor that depends on the field orientation. This situation is sketched in the lower part of Fig. 4e. Electrons at $E_F$ now encounter this gap at the weak points, and either backscatter or pass by thermal activation introducing a factor $e^{-\Delta/k_B T}$ in the transmission probability. Comparing this with the approximate $e^{-\alpha B_\parallel/T}$ behavior of $G_{edge}$ in Fig. 3c gives $g = \alpha k_B/\mu_B \approx 7.5$.

The insulating behavior in the 2D bulk of monolayers below ~100 K (see Fig. 2d and Supplementary Information 5) could be interesting because electron-hole correlations, as in an excitonic insulator[35,36], may alter the spectrum in a 2D conductor with either small band overlap or a small band gap. Unfortunately, it cannot be studied by standard techniques because of the edge conduction. By analyzing the variation with contact spacing in device MW3 above 150 K we were able to extract an approximate resistivity per square, and a contact resistance to the bulk of ~2 kΩ (see Supplementary Information 4). However, at low temperatures the contact resistance becomes too high. These difficulties may be overcome in the future by doping the contacts and by finding a way to make a Corbino geometry.

If monolayer WTe2 is indeed a monolayer 2DTI, with helical edge conduction at accessible temperatures as high as 100 K, it will afford new opportunities in the realms of topological and low-dimensional science. On the one hand, unlike the electrostatic confinement in quantum wells the edge of an exfoliated monolayer is abrupt, and its orientation, roughness, and chemical details are likely to be relevant especially to mesoscopic effects. Control of these factors may be possible by passivation or epitaxial growth[37]. On the other hand, tuning the band structure is possible by doping during growth or applying strain, and the electronic properties can be probed by direct surface techniques such as scanning tunneling spectroscopy. As a monolayer it can also be combined with layered magnets, semiconductors, and superconductors, for example to manipulate spin polarization or to create Majorana modes.[3,38]

## Acknowledgements

We thank Di Xiao, Anton Andreev and Boris Spivak for discussions. This work was supported by the U.S. Department of Energy, Office of Basic Energy Sciences, Division of Materials Sciences and Engineering, Awards DE-SC0002197 (DHC) and DE-SC0012509 (XX); AFOSR FA9550-14-1-0277; and NSF EFRI 2DARE 1433496.

# Supplementary Information

**Contents**
SI-1. Preparation and characterization of monolayer WTe$_2$ devices
SI-2. Magnetoresistance and temperature dependence of thick WTe$_2$ on SiO$_2$
SI-3. Measurements demonstrating that the edges are strongly coupled to the contacts
SI-4. Length dependence of the monolayer 2D bulk conduction
SI-5. Temperature dependence of the monolayer 2D bulk conduction
SI-6. Linear response measurements

**SI-1. Preparation and characterization of monolayer WTe$_2$ devices**

hBN crystals were mechanically exfoliated onto substrates consisting of 285 nm thermal SiO$_2$ on highly p-doped silicon under ambient conditions. 14-30 nm thick hBN flakes were chosen for the lower dielectric, and 5-12 nm thick flakes were used for the upper dielectric. Pt or Pd metal contacts (no substantial difference was found for these two metals) were deposited (5-7 nm) on the lower hBN by standard e-beam lithography and metallized in an e-beam evaporator. The upper hBN was picked up using polymer based dry transfer technique[1] and then moved into an oxygen-free glove box together with the lower hBN/contacts. WTe$_2$ crystals were exfoliated inside the glove b1ox and a monolayer flake was optically identified and quickly picked up with the upper hBN; the stack was then completed by transferring onto the lower hBN/contacts. Thus the WTe$_2$ was fully encapsulated before taking out of the glove box. After dissolving the polymer, a few-layer graphene (3-5 nm thick) was transferred onto the BN/WTe$_2$/BN stack as the top gate (except for device MW2, in which the top graphite (few-layer graphene) was transferred after the last metallization process). Finally, another step of e-beam lithography and metallization was used to define electrical bonding pads (Au/V) connecting to the metal contacts and the top gate.

Table S1 is a list of the thicknesses of upper hBN, lower hBN and top graphite flakes used in the reported devices. In all cases, we only biased the top gate to $V_g$, while the substrate was always grounded. Assuming the electron/hole density of states is not too small, the change in electron-hole density imbalance simply depends on the capacitance, $\Delta(n-p) = C_{tg}\Delta V_g/e$. Here $C_{tg}$ is the areal capacitance corresponding to the top gate, $C_{tg} = \epsilon_r\epsilon_0/d_{hBN}$, where $\epsilon_r \approx 4$ for hBN[2], $d_{hBN}$ is the thickness of the upper hBN flake.

| Device label | upper hBN (nm) | lower hBN (nm) | top graphite (nm) |
|---|---|---|---|
| MW1 | 5.8 | 18 | 3 |
| MW2 | 9.2 | 17.5 | 4.1 |
| MW3 | 11.4 | 14 | 3.4 |
| BW1 | 5.4 | 30 | 3.3 |
| TW1 | 5.5 | 22.6 | 4 |

**Table S1** | Thickness of the upper hBN, lower hBN, and the top graphite used for device MW1, MW2, MW3, BW1(bilayer) and TW1(trilayer). All thicknesses were obtained from the AFM image.

Fig. S1a is the optical image of device MW2 without top gate, Fig. S1b is the schematic image of a double-side gated device. After encapsulation, atomic force microscopy (AFM) was used to confirm the thickness of WTe$_2$. Clean edges produced step heights of approximately 0.7 nm for monolayers. Figure S2 shows Raman spectra of a monolayer WTe$_2$ encapsulated by hBN performed at 5 K with a 532 nm excitation laser. The peak locations are consistent with previous reports[3,4] on monolayer WTe$_2$, with no signature of breathing or shear modes above 10 cm$^{-1}$.

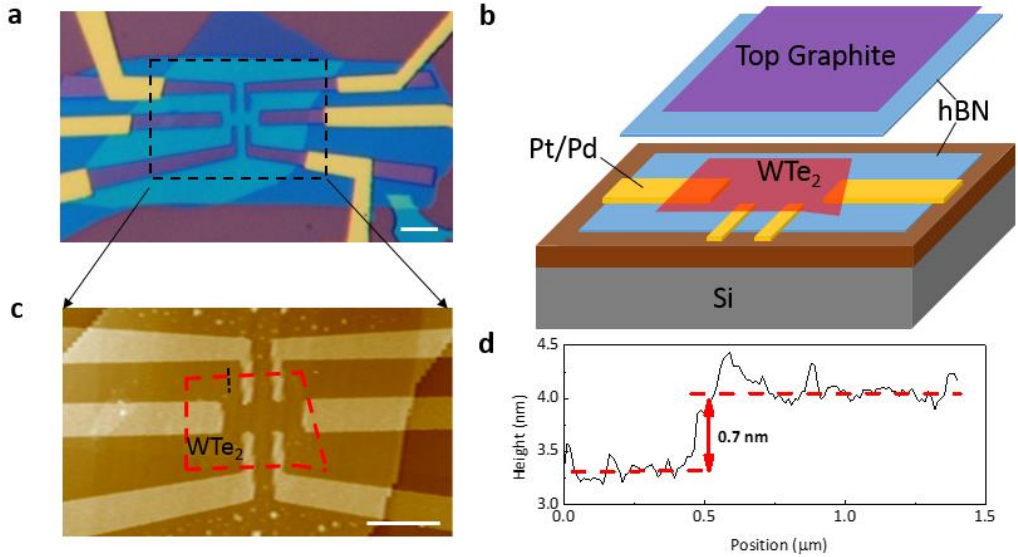

**Figure S1 | Monolayer WTe$_2$ device. a,** Optical microscope image of device MW2 without top gate. **b,** Cartoon of a typical monolayer WTe$_2$ device. The top graphite and hBN are show separated for clarity. **c,** Atomic force microscope image of the area highlighted in (a). The red dashed line outlines the monolayer flake. **d,** Line cut along the black dashed line in (c) matches the monolayer thickness, ~0.7 nm. Scale bars: 5 μm.

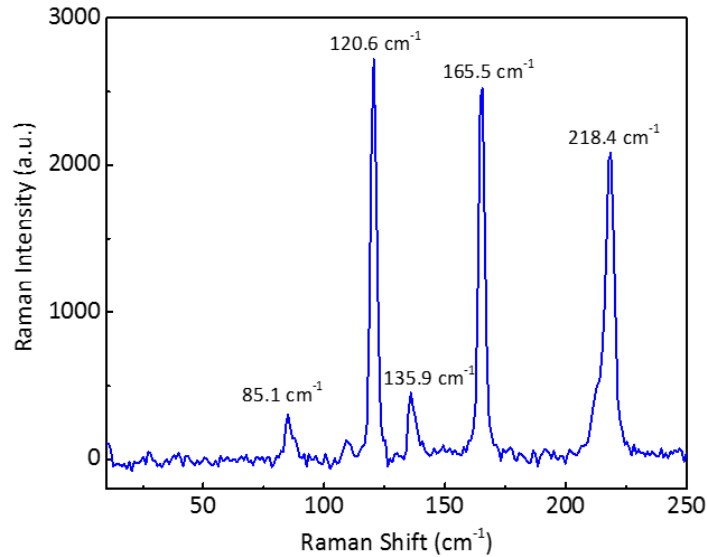

**Figure S2 | Raman spectra of an encapsulated monolayer WTe$_2$.** Five Raman peaks are clearly observed in the range 50 to 250 cm$^{-1}$, with wavenumbers 85.1 cm$^{-1}$, 120.6 cm$^{-1}$, 135.9 cm$^{-1}$, 165.5 cm$^{-1}$ and 218.4 cm$^{-1}$ respectively.

**SI-2. Magnetoresistance and temperature dependence of thick WTe$_2$ on SiO$_2$**

As is well established, monolayer WTe$_2$ degrades with time in air. However, for many-layer flakes only the top few layers appear to degrade, enabling electrical conductance measurements even without hBN encapsulation. Fig. S3a shows the four-terminal magnetoresistance (MR) as a function of magnetic field parallel to the c-axis measured on a 100 nm thick WTe$_2$ device with current along the a-axis. The large MR, 80,000%, at 7 T and 7 K is comparable to the previous report[5]. Fig. S3b shows a clear metal to insulator transition as layer thickness decreases for non-encapsulated flakes, consistent with the previous report[4]. As the temperature of the trilayer device decreases the resistivity actually increases, different from the encapsulated trilayer device reported in the main text (Fig. 1e). We also found non-encapsulated monolayer and bilayer flakes were not stable over time. In contrast, the encapsulated monolayer WTe$_2$ devices we measured in the main text are very stable, yielding indistinguishable results even after several months.

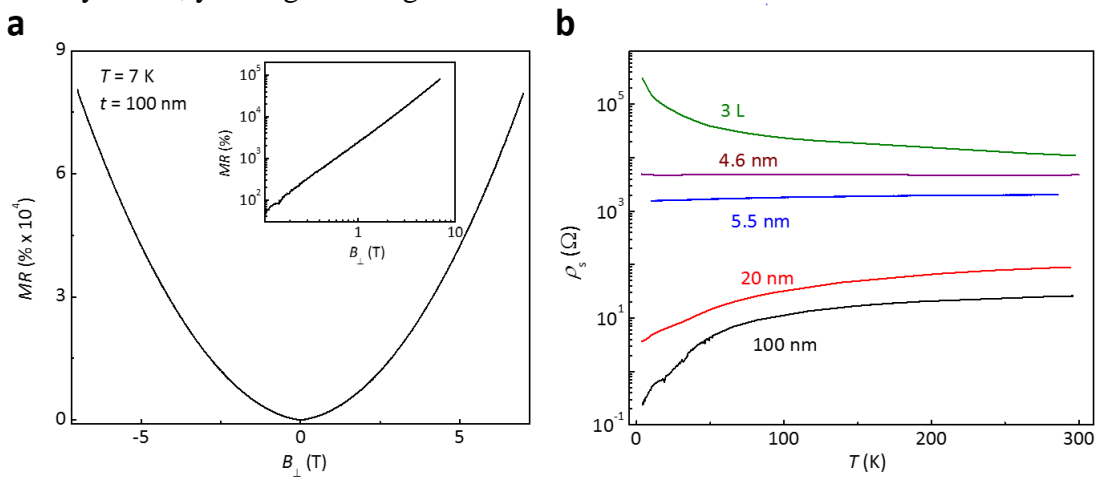

**Figure S3 | Bulk and few-layer WTe$_2$ on SiO$_2$ substrate. a,** Field dependence of MR in a 100 nm thick WTe$_2$ with the current along the a-axis and the applied field along c-axis at 7 K. The inset is the positive data on a log-log scale. **b,** Temperature dependence of sheet resistivity (per square) in non-encapsulated WTe$_2$ devices of different thickness, from 100 nm thick down to a trilayer.

### SI-3. Measurements demonstrating that the edges are strongly coupled to the contacts

In studying the edges in the main text (Figure 3 and 4), we focused on two-terminal measurements. Four-terminal measurements were not presented because they yielded the same results, implying the contacts to the edge were "perfect" in the sense that the current can only pass between adjacent edges via the metal of the contact in between. To illustrate this, Fig. S4a shows the parallel field dependence of the conductance at $V_g = 0$ in device MW2, in which the black and red curves were measured with two- and four-terminal configurations respectively as defined in the inset ($G_{23} = I_{23}/V_{23}$, $G_{14,23} = I_{14}/V_{23}$). Even as the magnetic field suppresses the edge conductance, both two and four-terminal measurements give almost identical conductance values. In Fig. S4b, we compare the gate dependence of $G_{13}$ with $(G_{12}^{-1} + G_{23}^{-1})^{-1}$ for three contacts in series in device MW3, where $G_{13}, G_{12}, G_{23}$ are two-terminal conductances. On the plateau region, there is almost no difference between the two, even for the mesoscopic fluctuations. This implies a strong contact coupling and, again, that the conductance is determined entirely by the edge in this regime, since bulk current flow between the pincer contacts would violate this equivalence.

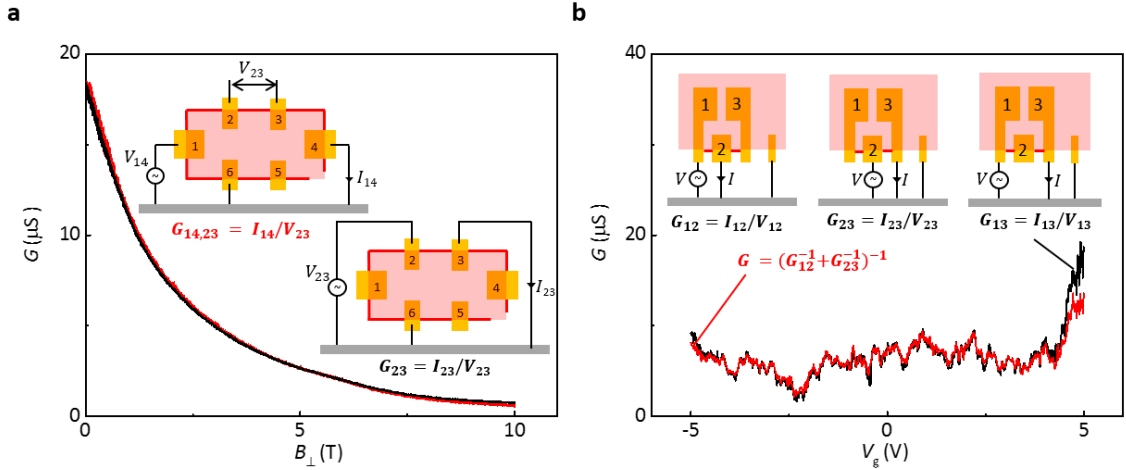

**Figure S4 | Comparison of two-terminal and multi-terminal measurements. a,** Perpendicular magnetic field dependence of the conductance of a particular edge measured in device MW2 at 6.5 K, $V_g = 0$; black and red curves are two- and four-terminal measurements respectively as labeled, with 2 and 3 the voltage contacts in both cases. **b,** Gate dependence of direct two-terminal conductance $G_{13}$ and of the series conductance $(G_{12}^{-1} + G_{23}^{-1})^{-1}$ in device MW3, at 1.6 K and $B = 0$.

## SI-4. Length dependence of the monolayer 2D bulk conduction

Above 100 K, two-terminal conduction is dominated by the 2D bulk. Fig. S5 shows the two-terminal resistance as a function of aspect ratio *L/W* for device MW1 at $V_g = 0$. If the edge contribution is small, the two-terminal resistance is given roughly by $R = \rho_s \frac{L}{W} + 2 R_c$. From the linear fit (for large aspect ratio a deviation from the linear fit is expected due to current spreading) we extract the sheet resistivity $\rho_s$, which increases from 20 kΩ at room temperature to 125 kΩ at 155 K, consistent with the insulating behavior for zero gate voltage. The extracted contact resistance $R_c$ is approximately 2 kΩ per contact.

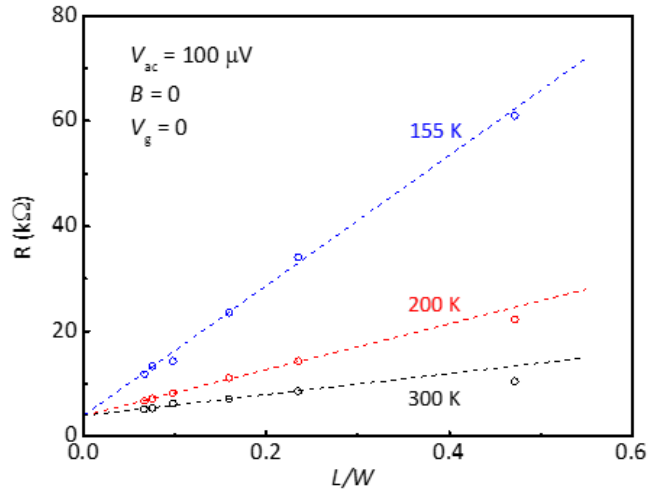

**Figure S5 | Length dependence of two-terminal resistance in device MW1.** Resistance as a function of aspect ratio at $V_g = 0$ for T = 300 K, 200 K, and 155 K.

## SI-5. Temperature dependence of the monolayer 2D bulk conduction

Figure S6a shows two-terminal linear-response conductance measurements in device MW2 made in a similar way to the measurements on the pincer-shaped device (MW3) in the main text (Fig. 2). Again a conductance plateau is seen at 4.2 K with mesoscopic fluctuations, indicating there is only edge conduction in this region. The conductance of the plateau is relatively low because of the combination of longer edges and ZBA. When we short out all the edge current, as shown in the insets, the conductance $I/V$ is suppressed to an unmeasurably small level in the plateau region implying that the conductance through the bulk is negligible at this temperature. Figure S6b shows its temperature dependence at $V_g = 0$. Above ~ 100 K, the conductance rises roughly linearly with temperature. At lower temperatures, it is approximately activated with activation energy ~5 meV (red trace in the right inset). This illustrates the sense in which the bulk becomes insulating below 100 K.

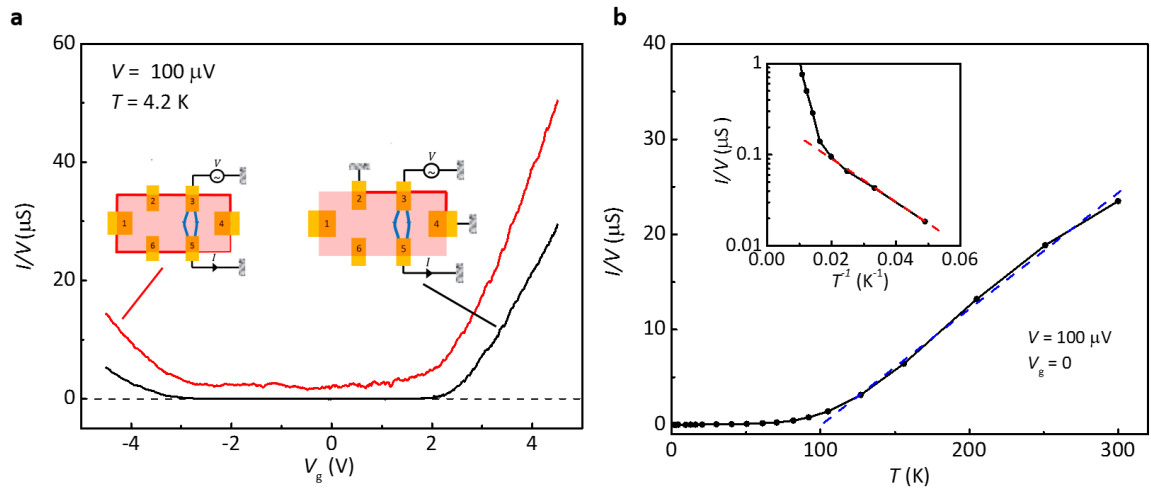

**Figure S6 | Gate and temperature dependence of bulk conductance in device MW2. a,** Comparison of *I/V* as a function of gate voltage for the two experimental configurations shown. Analogous to Fig. 2c in the main text, the red trace is the total two-terminal resistance which contains both edge and bulk contributions, while the black trace only contains a bulk contribution. **b,** Temperature dependence of the bulk measurement at $V_g = 0$. Inset: Arrhenius plot, showing approximately activated behavior below ~100 K (red trace, 5 meV). No signal was detectable above the background noise at temperatures below the lowest one shown here (20 K).

## SI-6. Linear response measurements

Fig. S7-10 are examples of transfer characteristics of different edges from base temperature 1.6 K to near room temperature in monolayer devices MW1, MW2, MW3 and bilayer device BW1 respectively. A small enough ac bias of 100 µV was used to ensure linear response. As can be seen from the temperature dependences, the strength of the ZBA varies for different devices and edges.

For the edge conductance values plotted in Fig. 4e in the main text, we used the averaged value over a gate range of (different for different edges) about 1 V near $V_g = 0$, where the plateau presents. In the case of the shortest edge of at 10 K in Fig. 4e, a bulk contribution of 2 µS was substracted.

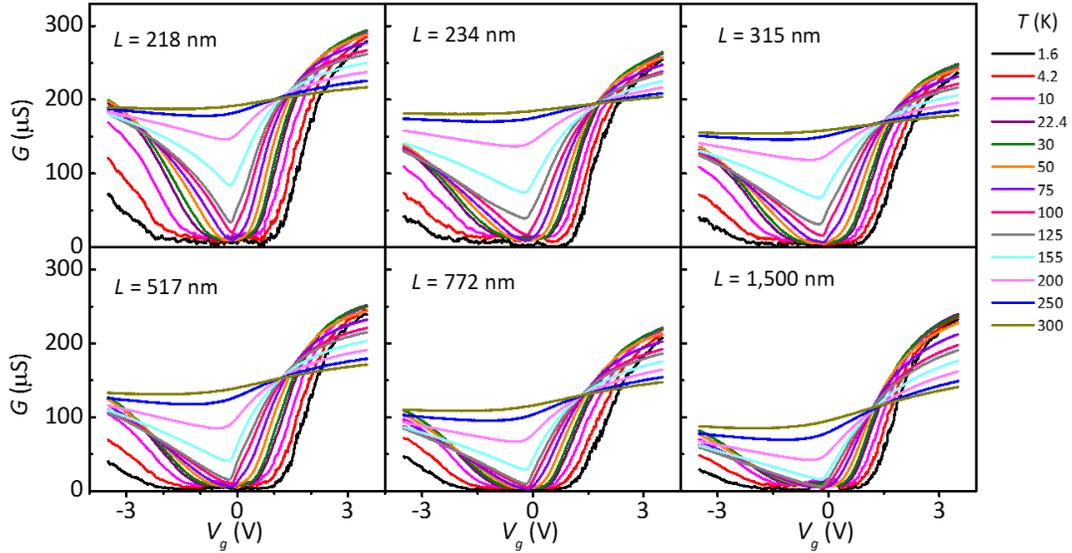

**Figure S7** | Temperature dependence of transfer characteristics of different edges (adjacent contact pairs) in monolayer device MW1, with contact separation $L$ ranging from 218 to 1,500 nm.

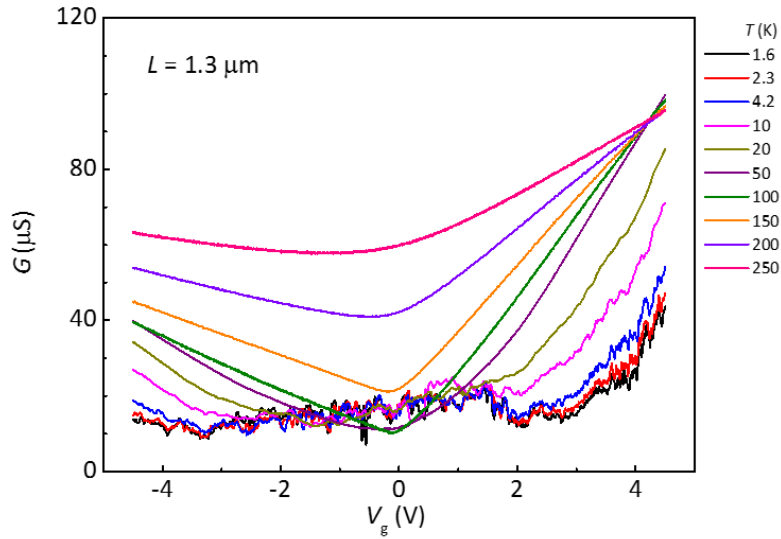

**Figure S8** | Temperature dependence of the transfer characteristic of a particular edge in monolayer device MW2. The contact separation is 1.4 µm.

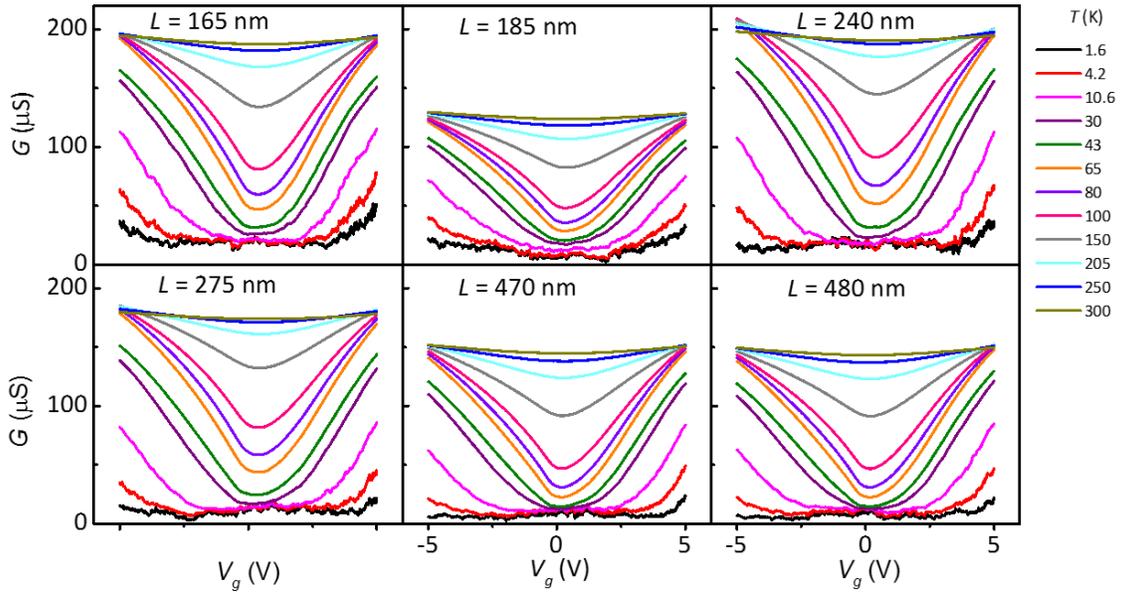

**Figure S9** | Temperature dependence of transfer characteristics of different edges in monolayer device MW3, with contact separation $L$ ranging from 165 to 480 nm.

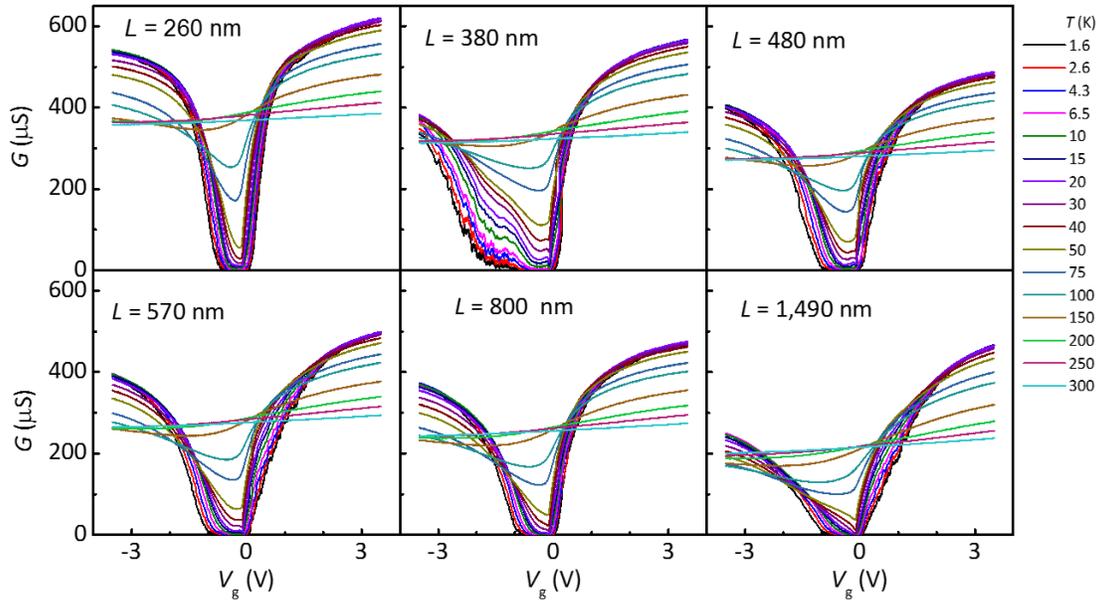

**Figure S10** | Temperature dependence of transfer characteristics of different edges in bilayer device BW1, with contact separation $L$ ranging from 260 to 1,490 nm.